\documentclass[times, twocolumn]{aastex63}

\newcommand{\vunit}{\mbox{\,km\,s$^{-1}$}}

\newcommand{\Msun}{\mbox{\,$M_\odot$}}
\newcommand{\Lsun}{\mbox{\,$L_\odot$}}

\newcommand{\mic}{\mbox{$\,\mu$m}}

\newcommand{\ltsimeq}{\raisebox{-0.6ex}{$\,\stackrel
	{\raisebox{-.2ex}{$\textstyle <$}}{\sim}\,$}}
\newcommand{\gtsimeq}{\raisebox{-0.6ex}{$\,\stackrel
	{\raisebox{-.2ex}{$\textstyle >$}}{\sim}\,$}}

\newcommand{\BminusV}{\mbox{$B$--$V$}}
\newcommand{\EBminusV}{\mbox{$E_{B-V}$}}

\newcommand{\pion}[2]{{#1}\,{\sc {#2}}}
\newcommand{\fion}[2]{[{#1}\,{\sc {#2}}]}
\newcommand{\nher}{\mbox{V1674~Her}}

\usepackage{graphics,graphicx}
\usepackage{epsfig}

\shorttitle{Infrared studies of nova V1674 Her}
\shortauthors{Woodward et al.}
\begin{document}
\title{Near-infrared studies of nova V1674 Herculis: A shocking record-breaker}

\author[0000-0001-6567-627X]{C. E. Woodward}
\affiliation{Minnesota Institute for Astrophysics, University of Minnesota,
116 Church Street SE, Minneapolis, MN 55455, USA}
\affiliation{Visiting Astronomer at the Infrared Telescope Facility, which is
operated by the University of Hawaii under contract NNH14CK55B\\
with the National Aeronautics and Space Administration.}

\author[0000-0002-9670-4824]{D. P. K. Banerjee}
\affiliation{Physical Research Laboratory, Navrangpura, Ahmedabad, Gujarat 380009, India}

\author[0000-0003-2824-3876]{T. R. Geballe}
\affiliation{NOIR Lab/Gemini Observatory, 670 N. A'ohoku Place, Hilo, HI 96720, USA}

\author[0000-0001-5624-2613]{K. L. Page}
\affiliation{School of Physics \& Astronomy, University of Leicester, Leicester LE1 7RH, UK}

\author[0000-0002-1359-6312]{S. Starrfield}
\affiliation{School of Earth \& Space Exploration, Arizona State University, 
Box 871404, Tempe, AZ 85287-1404, USA}

\author[0000-0003-1892-2751]{R. M. Wagner}
\affiliation{Department of Astronomy, Ohio State University, 140 W. 18th Avenue, Columbus, OH 43210, USA} 
\affiliation{Large Binocular Telescope Observatory, 933 North Cherry Avenue, 
Tucson, AZ 85721, USA}

\correspondingauthor{C.E. Woodward}
\email{chickw024@gmail.com}
\submitjournal{To Appear in The Astrophysical Journal Letters}

\begin{abstract}
We present near infrared spectroscopy of Nova Herculis 2021 (\nher),
obtained over the first 70~days of its evolution. This fastest nova on
record displays a rich emission line spectrum, including strong coronal
line emission with complex structures. The hydrogen line 
fluxes, combined with a distance of $4.7^{+1.3}_{-1.0}$~kpc, give an
upper limit to the hydrogen ejected mass of 
$M_{\rm ej}=1.4^{+0.8}_{-1.2}\times10^{-3} $\Msun.
The coronal lines appeared at day~11.5, the earliest onset yet
observed for any classical nova, before there was an obvious source of 
ionizing radiation. We argue that the gas cannot be photoionized, at 
least in the earliest phase, and must be shocked. Its temperature is 
estimated to be $10^{5.57 \pm 0.05}$~K on day~11.5. 
Tentative analysis indicates a solar abundance of aluminum and an 
underabundance of calcium, relative to silicon,  with respect to 
solar values in the ejecta. Further, we show that  the vexing problem of whether 
collisional or photoionization is responsible for coronal emission in classical novae 
can be resolved by correlating the temporal sequence in which the 
X-ray supersoft phase and the near-infrared coronal line  emission appear.
\end{abstract}


\section{Introduction}
\label{sec-Intro}

Classical nova (CN) systems consist of a semi-detached binary
containing a white dwarf (WD) and a Roche-lobe-filling
secondary star, usually a late-type dwarf. Material from the
secondary spills on to the surface of the WD through the inner
Lagrangian point, via an accretion disk. The material at
the base of the accreted envelope becomes degenerate, triggering a
thermonuclear runaway \citep[TNR; for a review see][]{2012clno.book.....B, 2020ApJ...895...70S}. 
Consequently, some $10^{-5}-10^{-4}$\Msun\ of material, enriched 
in C,N,O,Al,Mg following the TNR, is ejected explosively, 
at several 100 to several 1000\vunit.

CNe are characterized by the ``speed class'', the time $t_2$
($t_3$) taken for the visual light curve to decline by 2 (3)
magnitudes from maximum brightness \citep{warner2012}. There are
empirical relationships between speed class and the energetics
of the CN eruption: faster novae have the highest maximum bolometric
luminosities, largest outburst amplitudes, and fastest ejecta.
Indeed there is a relation between the absolute
magnitude at maximum and the rate of the light curve decline,
the ``MMRD'' relation \citep[see][and references therein]{2020A&ARv..28....3D}. 
These relationships are likely a consequence of the mass of the
WD on which the TNR occurred \citep{2020ApJ...895...70S}.

\section{Nova Her 2021 (\nher)}
\subsection{The nova}
Nova V1674~Her (TCP J18573095+1653396) was discovered by Seidji Ueda 
on 2021 June 12.5484 UT, at an apparent visual magnitude of 8.4;
it attained a peak brightness of $V \sim6$ mag \citep{2021ATel14704....1M, 2021RNAAS...5..160Q}. 
V1674 Her has been intensely observed since discovery, with optical 
spectroscopy at high cadence \citep[][and references therein]{2021ATel14723....1W}, near infrared (NIR) 
spectroscopy \citep[][and references therein]{2021ATel14765....1W}, and ultraviolet photometry and 
spectroscopy \citep{2021ATel14736....1K}. First detected in X-rays on June 14.41 UT,
V1674 Her began to show supersoft (SS) X-ray emission near July 1 \citep{2021ATel14747....1P}. 
\cite{drake21} give a detailed discussion of its early X-ray evolution. 
Radio observations reveal non-thermal emission \citep{2021ATel14758....1P, 2021ATel14731....1S}.
V1674 Her is also a $\gamma$-ray source \citep[][and references therein]{2021ATel14707....1L}.
NICER and Chandra X-ray observations \citep{2021ATel14798....1P, 2021ATel14776....1M}, as well
as pre-outburst archival $r$-band photometry \cite{2021ATel14720....1M}, reveal a 
$\sim$ 501~s oscillation period that is interpreted to be the spin period
of a WD in an intermediate polar system.

\nher\ is distinctive in many ways. Classification based optical spectra 
obtained within days of outburst \citep[][and references therein]{2021ATel14723....1W} indicate 
that \nher\ is of the FeII class \citep{1992AJ....104..725W}, with several FeII lines appearing, 
most prominent of which was the FeII(42) multiplet (however, the possibility
of a hydrid FeIIb classification based on a FeII to He/N transition around
$\sim$day 5.5 may be noted, \citet{2021ATel14728....1W}). At this early epoch the peak of the 
H$\alpha$ profile was corrugated, comprising around 9 to 10 distinct sub-peaks separated 
from each other in velocity by amounts ranging from 180 to 770~km~s$^{-1}$, while 
exhibiting an overall FWHM $\simeq 5850$~km~s$^{-1}$ The OI 8446~\AA\, line, 
dominantly excited by Lyman-$\beta$ fluorescence, was also prominent, having a 
FWHM of 5650~km~s$^{-1}$.  \nher\ also showed coronal line emission remarkably 
early. The detection of coronal lines at 11~days past maximum light is possibly the earliest reported 
detection of such lines in a nova \citep[see][]{1990AJ....100.1588B}. 

On day 28, \citet{2021ATel14746....1W} report the detection of broad 
[Ne V] 3426~\AA, as well as [Ne IV] at 4714~\AA, and 4724~\AA{} arguing that the 
presence of these strong neon emission lines is likely attributable to overabundances of 
neon. If substantiated, \nher{} would be a member of the ``neon nova'' class that includes 
QU Vul, V838 Her, and V1974 Cyg among others.

Here we discuss the exceptionally rapid near-infrared (NIR) 
spectral evolution of V1674 Her during the first $\sim70$~days post 
eruption, focusing our analyses on the very strong coronal emission 
which provides direct measures of gas temperatures, ejecta abundances,
and insight into the TNR process in novae. 

\subsection{Light curve}

\begin{figure}[ht!]
\includegraphics[width=0.45\textwidth]{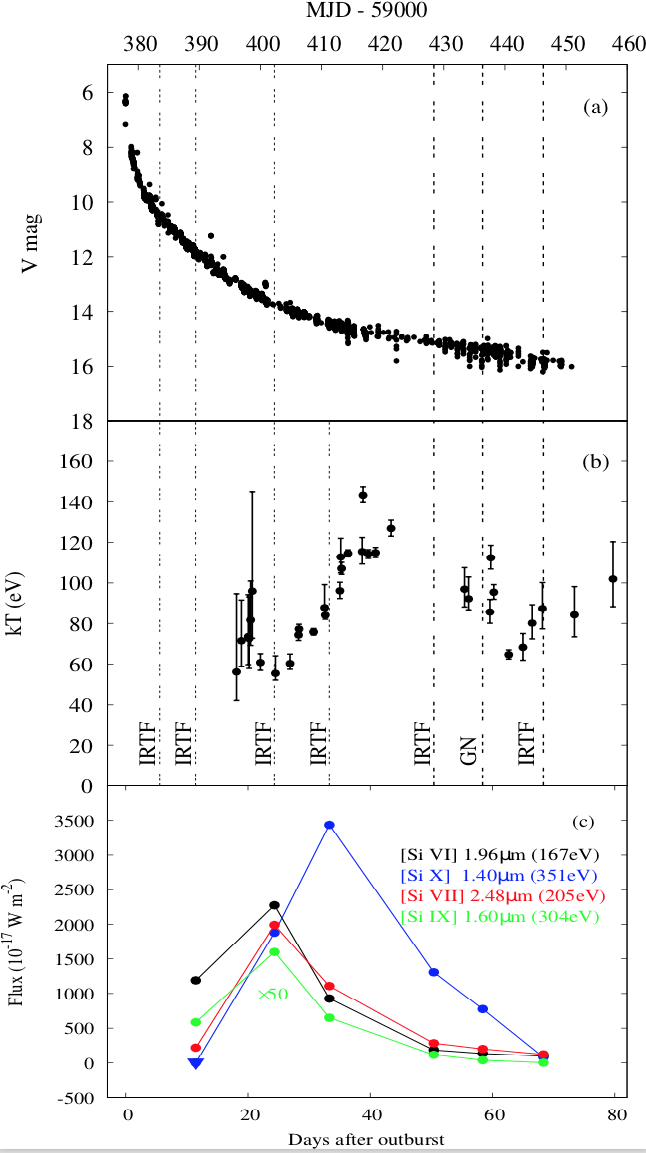}
\caption{(a): AAVSO $V$ band light curve with the dates 
(see Table~\ref{table-obs}) of the NIR spectroscopy marked.
(b): Temperature of the SS X-ray source as determined by fitting a 
blackbody to Swift data \citep[][]{drake21}. (c): Time-dependence of 
selected coronal line fluxes. Energies are IPs from Table~\ref{coronals}. 
Data for \fion{Si}{ix} 1.60~\mic\ have been multiplied by 50. See 
text for details.
\label{aavso}}
\end{figure}

From the AAVSO database \citep{kafka2021}, peak brightness was reached on 
MJD 59377.9605 (UTC 2021 June 12.96) at $V=6.14$, in agreement
with \cite{2021ATel14704....1M} who find $V = 6.18$ on Jun 12.90 UT.
We adopt Jun 12.96 UT (MJD~59377.96) as our time origin, 
$t_0$, and $V=6.14$ at maximum. Based on this $t_0$, and the AAVSO light 
curve in Fig.~\ref{aavso}, $t_2$, $t_3$ and $t_6$ are 
1~day, 2.2~days and 14~days respectively. 

When the $t_2$ and $t_3$ times are considered together, V1674~Her is again
remarkable. It is the fastest Galactic nova known to date,
of all the compiled, detailed light curves \citep{2010AJ....140...34S} surpassing V838~Her 
($t_2 = 1.0, t_3 = 3.0$) and the recurrent nova U Sco ($t_2 = 1.0, t_3 = 4.0$). 
Optical \citep{2021ATel14723....1W, 2021ATel14746....1W} and NIR \citep{2021ATel14765....1W} 
spectroscopy bear this out, with emission line widths and P~Cygni 
profiles indicating ejection velocities $\sim6000$\vunit. 
Such extreme eruptions are characteristic of a TNR that occurred 
on a WD with a mass very close to the Chandrasekhar limit \citep{2020ApJ...895...70S}. 

\subsection{Reddening and distance estimate}

The reddening, \EBminusV, is in the range 0.52--0.58 based on 
the 6613\AA\ Diffuse Interstellar Band feature, clearly seen in the 
high-resolution ($R=220$K) optical spectra \citep{2021ATel14723....1W}. 
From these high-resolution spectra, the interstellar 
resonance \pion{K}{i} line at 7699~\AA\ is measured to have an 
equivalent width of 0.1349~\AA, which gives a
reddening \EBminusV\ = 0.52 using the calibration by \cite{1997A&A...318..269M}.
\cite{2021ATel14704....1M}, from the \pion{K}{i} 7699~\AA\ line, measure 
\EBminusV\ = 0.55. The observed color of \BminusV = 0.72 and 0.55 at $t_0$ 
and $t_2$ respectively, when compared with the intrinsic colors
of novae expected at these early evolutionary 
times \citep{1987A&AS...70..125V}, yields \EBminusV values in the range 
0.49 to 0.57. We adopt \EBminusV\ = 0.55.

Adopting the method described in \citet[][and references therein]{2018MNRAS.473.1895B}, 
an independent estimate of the reddening, together with a distance
estimate, can simultaneously be obtained from the 
intersection of the extinction versus distance curve in the direction 
to the nova \citep{2006A&A...453..635M}, and the curve defined by 

\begin{equation}
 m_{\rm v} - M_{\rm V} = 5 \log{d} -5 + A_{\rm V} \label{mM}
 \end{equation}
 
\noindent where $m_{\rm v} = 6.14$ at maximum and $M_{\rm V} = -9.05\pm0.5$  
\citep[for $t_2=1$~d derived from the MMRD relation of][]{1995ApJ...452..704D}. 
The intersection of these two curves yields a distance of 
$4.7^{+1.3}_{-1.0}$~kpc and visual extinction $A_V = 1.82\pm0.33$, 
or \EBminusV\ = 0.59$\pm0.11$,  assuming that the total-to-selective 
extinction is 3.10. As a cross-check, applying this method to a 
similarly fast nova V838 Her ($t_2 =2$~d), gives  distance and 
$A_V$ estimates of 3.83~kpc and 1.40 magnitudes respectively, consistent
with other compiled measurements \citep{1994ApJ...437..827H}.  
Gaia estimates for the distance to \nher{}
\citep{2021AJ....161..147B, 2018AJ....156...58B} have large uncertainties (the 
parallax measurements are at the limit of usefulness), but the central 
values are consistent with the latter $4.7^{+1.3}_{-1.0}$~kpc estimate. 

The spectral energy distribution (SED) at maximum, constructed 
using $BV\!RI$ values \citep{2021ATel14704....1M}  dereddened by \EBminusV\  = 0.55, yield 
a blackbody fit with T $= 9900$~K. The outburst luminosity, 
assuming $d = 4.7$~kpc, is $L_{\rm out} = 3.7\times10^5$\Lsun, 
which is $\sim9$ times the Eddington luminosity of a 1.3\Msun\ WD.

\section{Observations and Data Reduction}

\begin{deluxetable}{@{\extracolsep{0pt}}lcccccl}
\tablenum{1}
\setlength{\tabcolsep}{3pt} 
%
%
\tablecaption{Observational Log V1674 Her\tablenotemark{$\dagger$}\label{tab:obstab}}
\tablehead{
& \colhead{Mean} &  &\colhead{SpeX} & &\colhead{Average}  & \\
\colhead{Date} &\colhead{MJD} & Day\tablenotemark{a} &\colhead{Mode\tablenotemark{b} } &\colhead{IT\tablenotemark{c} } &\colhead{Airmass} &\colhead{Standard }\\  
\colhead{(2021 UT)}  & (-59000) && &\colhead{(sec)}  
 }
\startdata 
Jun 18.54 & 383.544 & 5.64  & SXD    &  356 & 1.08 & HD165029\\
Jun 24.42 & 389.409 & 11.51 & SXD   &  356 & 1.06 & HD165029\\
Jun 24.42 & 389.424 & 11.52 & LXD   &  445 &  1.03 & \nodata \\
Jul 07.27  & 402.274 & 24.37 & SXD   &  657 & 1.65 & HD165029\\
Jul 07.29  & 402.291 & 24.39  & LXD  & 834  & 1.47 & \nodata \\
Jul 16.27  & 411.268 & 33.37 & SXD   & 956 & 1.46 & HD177724\\
Aug 2.34  & 428.341 & 50.38 & SXD   & 1977 & 1.01 & HD165029\\
Aug 2.37  & 428.373 & 50.47 & LXD   &  890 & 1.01 &  \nodata\\
Aug 10.35 & 436.353 & 58.39 & GNIRS\tablenotemark{d} & 720 & 1.01 & HIP91118 \\
Aug 20.28 & 446.284 & 68.32 &SXD   &  478 & 1.01 &  HD165029\\
\enddata
\tablecomments{$^{\dagger}$Reduced spectra are provided as meta-data behind the figures.\, 
$^{a}$Day $t_o$ =  Jun 12.96 UT (MJD~59377.96).\, $^{b}$All SpeX spectra were obtained 
with a 0\farcs5 $\times$ 15\farcs0 slit (width $\times$ length), yielding an effective 
spectral resolution $R=1200$. The slit was set to the parallactic angle at the time of 
observations.\, $^{c}$\,Total on-source integration time on the nova.\, $^{d}$Observed with 
Gemini North / GNIRS SXD, 0\farcs45 slit $R=1200$.}
\label{table-obs}
\end{deluxetable}

\begin{figure*}[!ht]
\figurenum{2}
\begin{center}
\gridline{
\rotatefig{0}{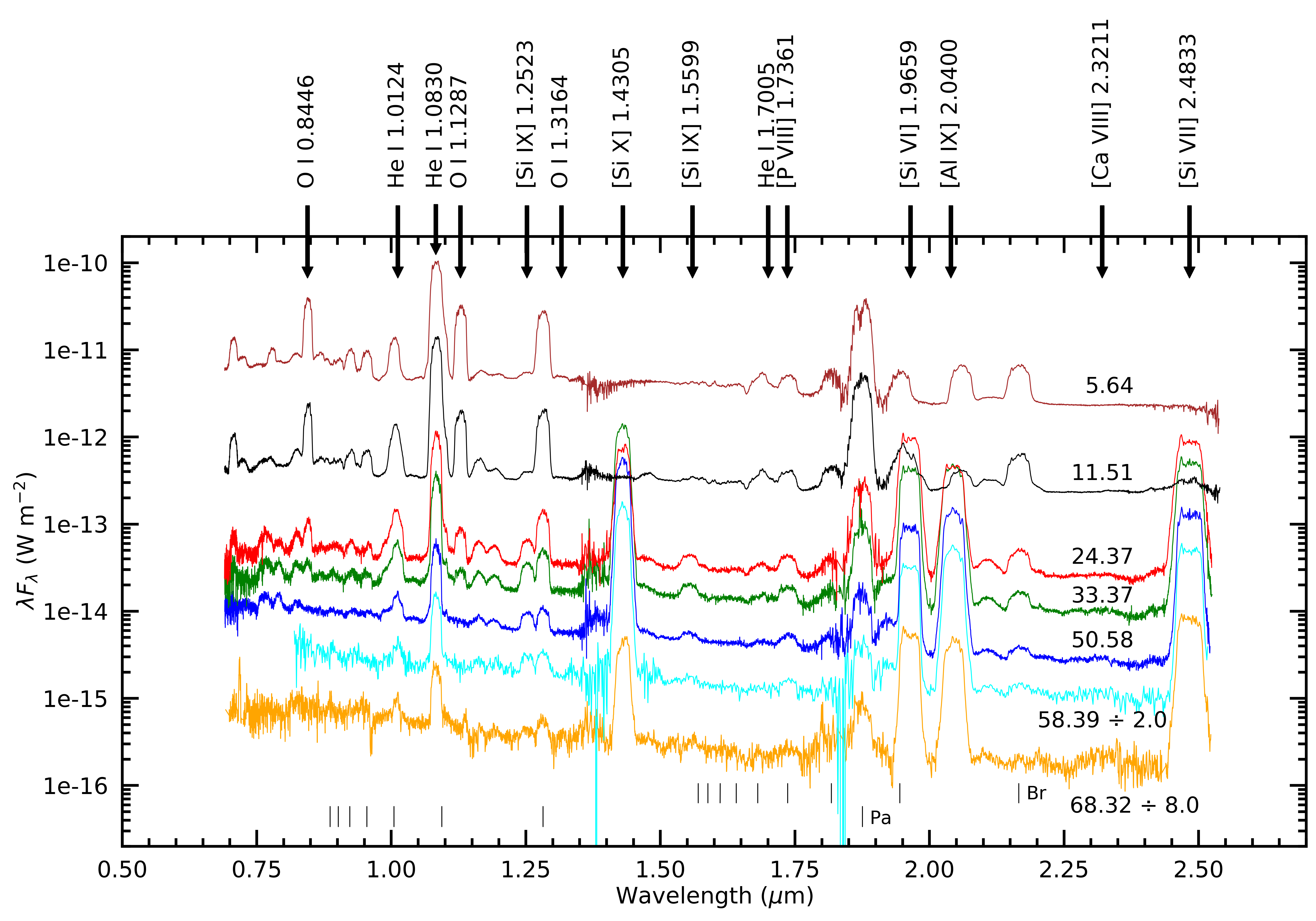}{1.15\columnwidth}{(a)} 
}
\gridline{
\vspace{-1.1cm}\rotatefig{0}{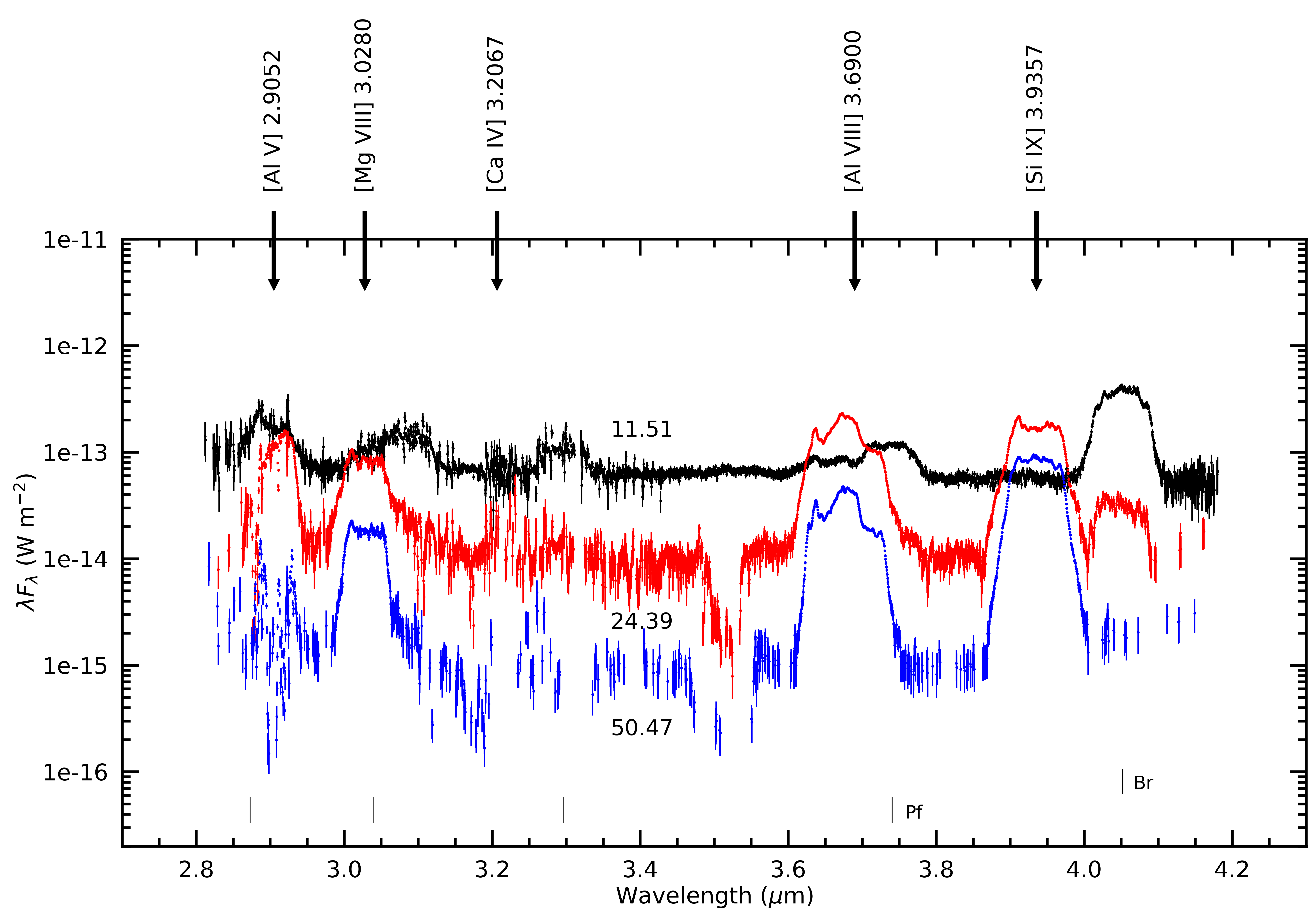}{1.15\columnwidth}{(b)} 
}
\vspace{1.0cm}\caption{\label{fig-flick} 
\small{Observed near-infrared spectra of \nher\ at different epochs (Table~\ref{table-obs}).
The spectral points are clipped with a SNR threshold of $\gtsimeq 2.5$. Coronal lines 
(see Table~\ref{coronals}), and representative He and O lines are identified. The small 
ticks in the bottom of each panel indicate the wavelength of various hydrogen emission line series
(Brackett, Paschen, Pfund). (a) 0.7 to 2.5\mic\ range; data for day~58.39 and 68.32 have been divided by factors 
of 2 and 8 respectively, for clarity. 
(b) 2.7 to 4.3\mic{} range (SpeX LXD mode thermal wavelengths). The spectrum on day~68.32 in panel (a)
and spectra on days~24.39 and 50.42 in panel (b) have been smoothed using a Savitzky-Golay filter
using a SG window=7.5 and degree=2 \citep[for a description see][]{2004PASP..116..362C}. 
(The observed spectra are available as data behind the figures). }
}
\end{center}
\end{figure*}

Spectra of V1674 Her were obtained with SpeX \citep{2003PASP..115..362R} 
on the 3.2~m NASA IRTF. These data were reduced with corrections for 
telluric absorption(s) using the SpexTool pipeline \citep{2004PASP..116..362C}. 
The spectra were flux calibrated using an A0 standard star 
observed at a comparable airmass $\Delta_{AM} \ltsimeq 0.14$. 
The IRTF flux calibrations are accurate to $\pm 10$\%.
The observational log is presented in Table~\ref{table-obs} 
and the spectra are shown in Fig.~\ref{fig-flick}.
The times of the observations in relation to the visual
light curve are shown in the top panel of Fig.~\ref{aavso}.

$JH\!K$ photometry also was obtained with the SpeX guider imager using
a 10 point dither with 14.93~s exposure time at each position, and at an
average air mass of 1.13. Data reduction used standard infrared 
techniques. Instrumental magnitudes obtained through aperture photometry 
were calibrated against several 2MASS field stars. The $JH\!K$ magnitudes were
$J=12.74\pm0.10, H=12.48\pm0.12, K=10.58\pm0.05$ (2021 July 16.33)
and $J=13.81\pm0.05,H=13.59\pm0.03, K=11.59\pm0.14$ (2021 Aug 2.44).

A single 0.82-2.52\mic{} spectrum of V1674 Her was obtained on UT 2021 August 10 
at the Frederick C. Gillett Gemini North 8~m Telescope, with the facility 
spectrometer GNIRS \citep{2006SPIE.6269E..4CE} for program GN-2021B-FT-101 (see Table 1). 
Data reduction utilized IRAF\footnote{IRAF is distributed by the National 
Optical Astronomy Observatories, \\ which are operated by the Association 
of Universities for Research in \\ Astronomy, Inc., under cooperative 
agreement with the National \\ Science Foundation.} and Figaro commands 
and produced the spectrum shown in Fig.~\ref{fig-flick}. The GNIRS observations 
were obtained in poor seeing and through thin but variable clouds; 
consequently the flux calibration is uncertain by $\sim30$\%.

\section{Results and Discussion}

The spectra are shown in Fig.~\ref{fig-flick}, the two panels 
covering the wavelength ranges 0.7--2.5\mic\ and 1.8--4.1\mic,
corresponding to the SpeX Short-Cross Dispersed (SXD) and Long-Cross Dispersed
(LXD) wavelength ranges, respectively. Here, the discussion focuses mainly on the 
coronal lines; a more detailed discussion of the SEDs is deferred to a later paper.

\subsection{Ejecta mass estimate}

Using the dereddened fluxes of Pa-$\beta$ and Br-$\gamma$, 
we ascertain that their relative strengths follow Case~B predictions
on day~11.51, within the parameter space expected to represent 
the ejecta, viz.  $T = 5000$~K to 20000~K, and electron density
$n_e \sim 10^6$ to $10^9$~cm$^{-3}$. Other H lines could not be 
used because of blending with neighboring lines, leading to 
inaccurate line flux estimates. We assume the ejecta density
on day~11.51 to be $n_e = 10^{6.6}$~cm$^{-3}$ since the \fion{Al}{vi} 3.66\mic\ line, 
which has a critical density of $n_e = 10^{6.6}$~cm$^{-3}$ at $T = 10000$~K 
\citep{1993ApJS...88...23G}, is detected on this day. 

The ejecta mass is given by

\begin{equation}
\frac{M_{\rm ej}}{\Msun} = 2.22\times10^{-9} \left( \frac{d}{4.7~kpc} \right)^2  \frac{f}{n_e\epsilon}
\end{equation}

\noindent where $f$ is the dereddened flux in an H line in W~m$^{-2}$, 
$\epsilon$ is the emissivity in the line in erg~s$^{-1}$~cm$^{3}$ 
\citep{1995MNRAS.272...41S}, and $n_e$ is the electron density per cm$^{3}$. Using the Pa-$\beta$ 
and Br-$\gamma$ lines, and assuming $d = 4.7^{+1.3}_{-1.0}$~kpc and 
a pure H composition, the ejecta mass is 
$M_{\rm ej} = 1.4^{+0.8}_{-1.2}\times10^{-3} \times(10^{6.6}/n_e)$\, \Msun, wherein
the upper limit on the mass is obtained by choosing $n_{e} = 10^{6.6}~\rm{cm}^{-3}$.

%

\begin{deluxetable*}{@{\extracolsep{0pt}}llcccccccccccc}
\tablenum{2}
\setlength{\tabcolsep}{4pt} 
\tablecaption{V1674 Her Coronal line dereddended fluxes\label{coronals}}
\tablehead{
 \colhead{Ion} & \colhead{$\lambda$} &  & \colhead{Transition}&  & \colhead{IP} & \colhead{$kT_{*}$}
  & \colhead{$\Psi$} & \multicolumn{6}{c}{Line fluxes ($\times 10^{-17}$~W~m$^{-2}$)}\\ 
      &  \colhead{($\mu$m)}  & & \colhead{$u-\ell$} &&\colhead{(eV)}& \colhead{(eV)}  && \colhead{11.51d} & \colhead{24.38d} & \colhead{33.37d}
       & \colhead{50.43d}   & \colhead{58.39d} & \colhead{68.32d}
       }  
\startdata 
\fion{S}{ix}\tablenotemark{a}    & 1.2523  && $^3$P$_1-^3$P$_2$&& 329 & 137 & $\ldots$ & 1736 & 98 & 49 & 9.2 & 5.0 & 1.7\\ 
\fion{Si}{x}    & 1.4305  && $^2$P$^o_{3/2}-^2$P$^o_{1/2}$&& 351 & 146 & $\ldots$ &  $<2.3$ &1878 & 3437 & 1307 & 774 & 82 \\ 
\fion{Si}{ix}   & 1.5599  && $^3$P$_2-^3$P$_0$ && 304 & 127 & $\ldots$ & 11.7 & 32 & 13 & 2.3 & 0.7 & $<$\\ 
\fion{P}{viii}\tablenotemark{b}  & 1.7361  && $^3$P$_1-^3$P$_2$ && 264 & 110 & $\ldots$ &  231 & 32 & 15 & 2.1 & 1.4 & $<$\\
\fion{Si}{xi}   & 1.9320   && $^3$P$^o_2-^3$P$^o_1$&& 401 & 167 & $\ldots$ &  $<8.5$ & & & $<1.8$ & $<0.2$ & \\ 
\fion{Si}{vi}   & 1.9650  && $^2$P$^o_{1/2}-^2$P$^o_{3/2}$&& 167 & 70  & 0.423  & 1190 & 2274 & 927 & 185 & 133 & 95 \\ 
\fion{Al}{ix}\tablenotemark{c}   & 2.0400 && $^2$P$^o_{3/2}-^2$P$^o_{1/2}$ & & 285 & 119 & $\ldots$  & $<4$ & 1035 & 1004 & 296 & 216 & 72 \\
\fion{Ca}{viii} & 2.3211  && $^2$P$^o_{3/2}-^2$P$^o_{1/2}$ && 127 & 53 &  $<5.88$ & 5.1 & 1.25 & & $<0.6$ & $<0.1$ & \\ 
\fion{Si}{vii}  & 2.4833  && $^3$P$_1-^3$P$_2$&& 205 & 86 &  0.695 &  218 & 1990 & 1107 & 283 & 200  & 121 \\ 
\fion{Al}{v}    & 2.9052  && $^2$P$^o_{1/2}-^2$P$^o_{3/2}$ && 120 & 50 & 0.456 & 51 & 287 & --- &  & --- & --- \\ 
\fion{Mg}{viii} & 3.0280  && $^2$P$^o_{3/2}-^3$P$^o_{1/2}$&& 225 & 94 & $\ldots$ & $<5$ & 288 & --- & 103 & --- & --- \\ 
\fion{Ca}{iv}\tablenotemark{d} & 3.2067 && $^2$P$^o_{1/2}-^2$P$^o_{3/2}$&& 51 & 21 & $\ldots$ &&     & ---    & & --- &  --- \\ 
\fion{Al}{vi}   & 3.6597  && $^3$P$_1-^3$P$_2$ && 154 & 64 & 1.365 & 47 & &---   &  & --- & --- \\ 
\fion{Al}{viii} & 3.6900  && $^3$P$_2-^3$P$_1$ && 242 & 101 & $\ldots$  & & & --- & & --- & --- \\ 
\fion{Si}{ix}   & 3.9357  && $^3$P$_1-^3$P$_0$ && 304 & 127 & $\ldots$ & $<1.5$ & 614 & --- & 468 & --- & --- \\ 
\enddata
\tablecomments{Upper limits are $1\sigma$. IP is the ionization potential of the lower ionization state.
See text for explanation of $kT_{*}$. Table entry with ``---'' indicates that the observation did not cover the wavelength range.
$^{a}$Heavily blended with \pion{He}{i} 1.253 $\mu$m.\, $^{b}$Blended with \pion{H}{i} 10--4 1.737 $\mu$m.\,
$^{c}$Blended with \pion{He}{i} 2.058 $\mu$m.\, $^{d}$Not strictly coronal according to the definition of \cite{1990ApJ...352..307G}.
The effective collision strengths ($\Psi$), are the collisional strengths $\Omega$ averaged over a thermal electron 
distribution and are taken from the IRON Project \citep{2006IAUS..234..211B, 1993AA...279..298H} on-line
database (\url{http://cdsweb.u-strasbg.fr/tipbase/home.html}) interpolated to
a temperature of $10^{5.57 \pm 0.05}$. 
}
\end{deluxetable*}


\subsection{The coronal lines}

\begin{figure}[!ht]
\figurenum{3}
\begin{center}
\gridline{
\rotatefig{0}{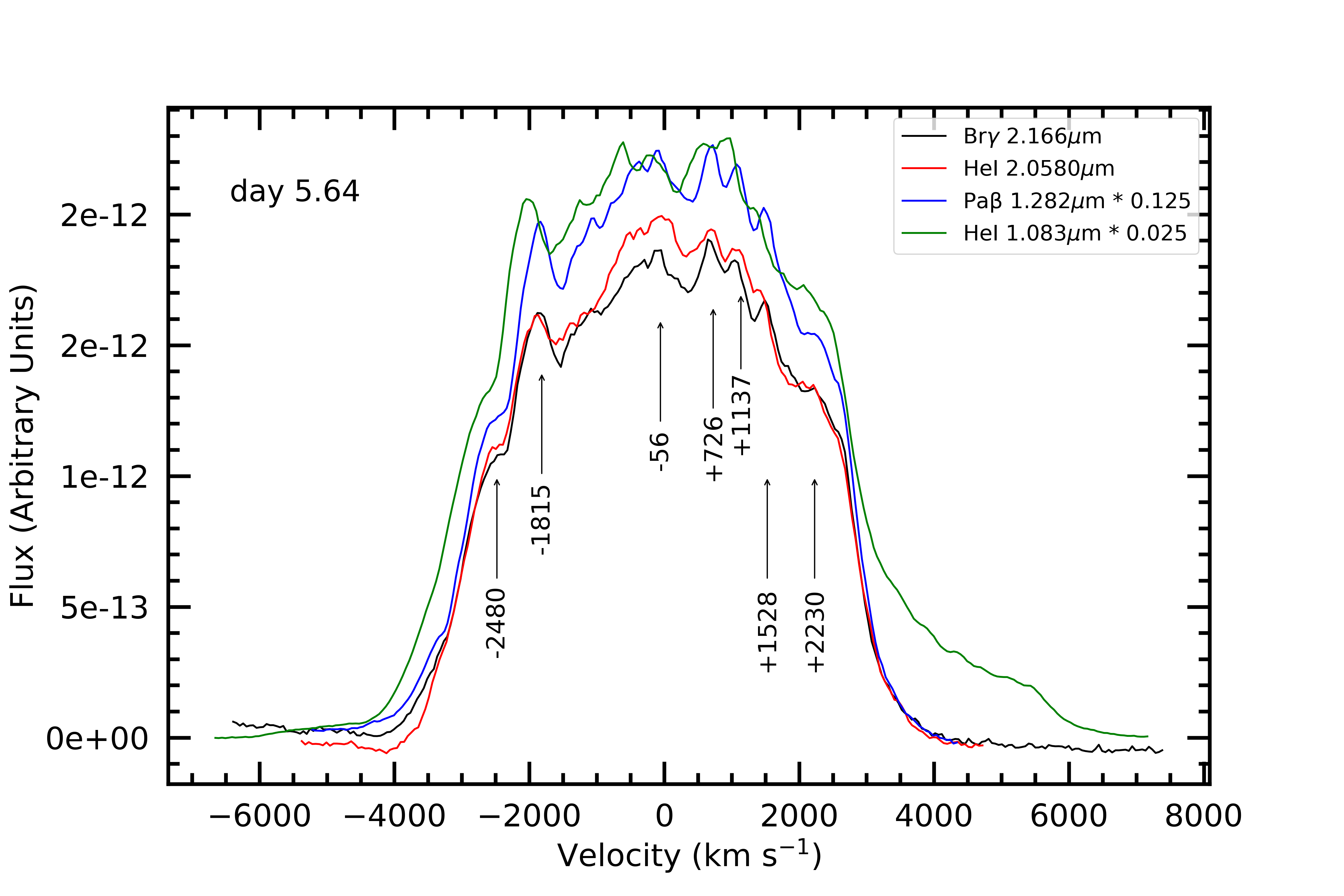}{1.00\columnwidth}{ }
}
\vspace{-1.00cm}
\gridline{
\rotatefig{0}{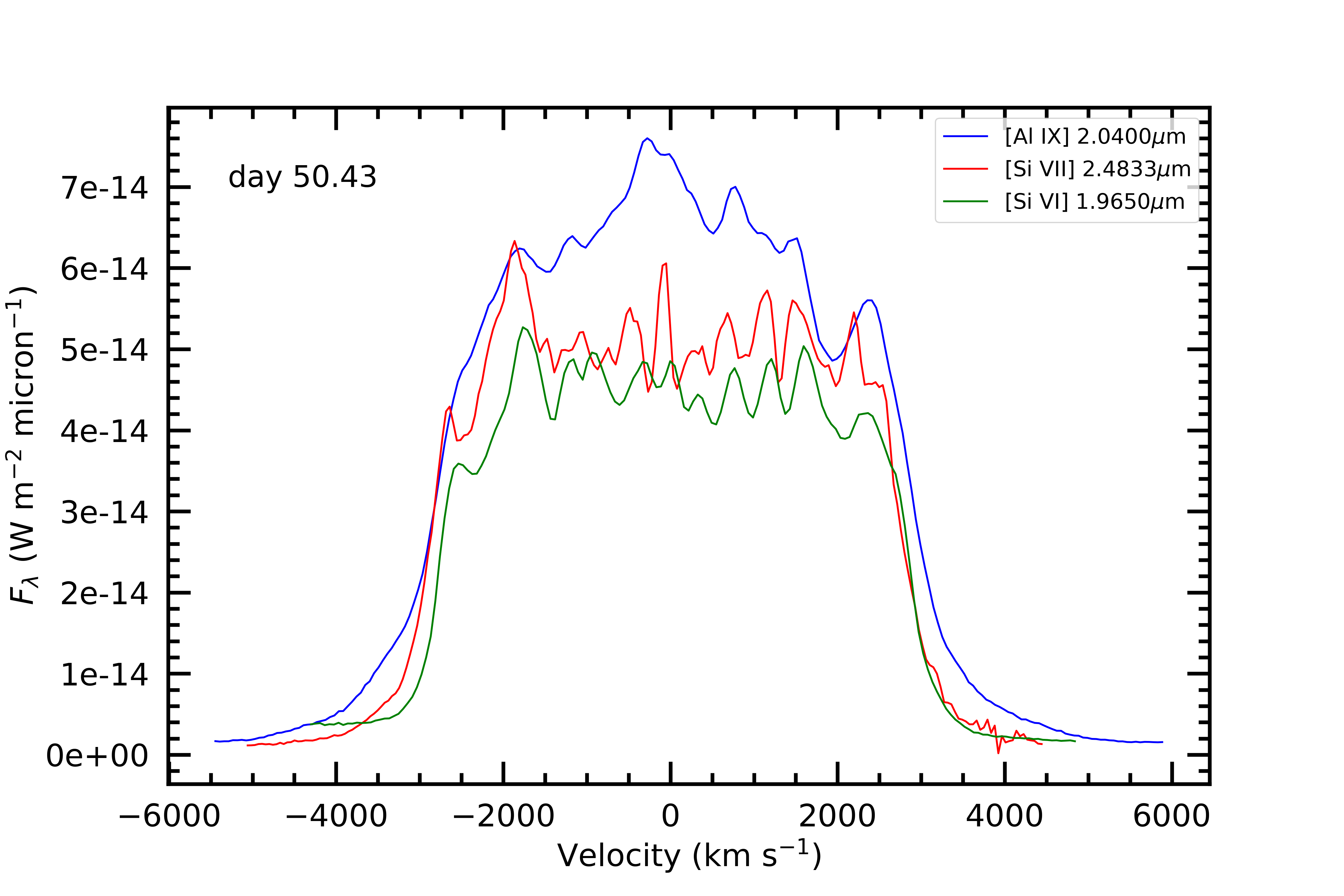}{1.00\columnwidth}{ }
}
\vspace{-0.75cm}
\caption{ \label{profiles} Velocity profiles of the emission lines. (a) Two H recombination lines 
and two \pion{He}{i} lines on day~5.64 in velocity space. (b) Coronal lines on day~50.43.}
\end{center}
\end{figure}

The profiles of four nebular lines (two \pion{H}{i} and two \pion{He}{i}) on day 5.64 are shown in 
Fig.~\ref{profiles}a to illustrate the complexities of the line profiles. The similarities of the 
details in each of these profiles are remarkable. In particular, there are identifiable features 
at radial velocities --2480\vunit, --1820\vunit, --60\vunit, 730\vunit, 1140\vunit, 1530\vunit\ 
and 2310\vunit, each with an uncertainty of $\pm100$\vunit. The velocity structure seen at an 
epoch (day 50.43) when the coronal lines are strong, Fig.~\ref{profiles}b, is similar.

The coronal line fluxes are listed in Table~\ref{coronals}. Uncertainties in the flux 
determinations are estimated to be $\ltsimeq \pm20$\% for the stronger lines, and 
as much as $\pm40$\% for the weaker. The table also lists the ionization potential 
(IP) of the lower ion (thus the entry for \fion{S}{ix} gives the IP  of \pion{S}{viii}). The 
parameter $kT_*$ denotes the temperature $T_*$ at which half the photons emitted by 
the corresponding blackbody can ionize the lower ion; we take this to imply the temperature 
of the ionizing source that is capable of producing the requisite ion.

The time-dependence of the silicon coronal lines, among the 
strongest and best recorded coronal lines in the spectrum, 
is shown in Fig.~\ref{aavso}. It is clear from Table~\ref{coronals}
and Fig.~\ref{aavso} that coronal line emission 
was first unambiguously detected, via the \fion{Si}{vi} 1.96\mic, 
\fion{Si}{vii} 2.48\mic, and the \fion{Al}{vi} 3.66\mic\ 
lines, on Jun 24 (day 11.51), well before the SS phase began 
\citep{2021ATel14747....1P, drake21}. In this respect \nher\ is again a record 
breaker. Both V1500 Cyg and V838 Her, two of the fastest novae 
with $t_{2}$ times of 2 and 1 day(s) respectively, were first detected to show coronal 
emission at 29 and 17 days, respectively, past maximum 
light \citep{1993JApA...14....7C, 1990AJ....100.1588B}. 

The presence of coronal lines so early in the evolution
implies that, despite the presence of a hot central WD
remnant, the initial coronal emission could not have been caused by 
photoionization. For example, \fion{Si}{vii} 2.48\mic\ is 
present on day 11.51 even though the ionization of \pion{Si}{vi}
requires 205~eV, far exceeding the temperature of the 
ionizing source ($kT_*=86$~eV) that would emit sufficient photons,
and is well in excess of the source temperature as measured
by The Neil Gehrels Swift Observatory (see Fig.~\ref{aavso}).

We therefore conclude that the coronal emission arose from  
collisional ionization by particles that gained sufficient
energy from shock heating. This contradicts the commonly assumed 
view that, when the central source is sufficiently hot,
photoionization is the principal mechanism for coronal line 
generation \citep[see][]{1990AJ....100.1588B}. 
Plausible support for shocks comes from the 
multi-peaked line profiles shown in Fig.~\ref{profiles}, 
which were even more clearly resolved into a large number 
of sub-structures in very high resolution ($R = 220,000$) spectra
\citep{2021ATel14723....1W}. This points to parcels of gas in the 
ejecta having widely different velocities. Collisions between such parcels would not 
only lead to shock  heating, but might also account for the gamma ray emission 
seen in the early stages of V1674 Her \citep{2021ATel14705....1L}.

To make estimates, we suppose that the ejecta initially are fully
ionized and have a temperature $\sim10^4$~K, typical for CNe. 
For a gas of solar composition, the root-mean-square ion 
speed is $V_{\rm rms}\sim20$\vunit.
Fig.~\ref{profiles} suggests that there must exist in the 
ejecta parcels of gas with relative speed $V_{\rm rel}\sim1000$\vunit.
Thus $V_{\rm rel}/V_{\rm rms}\sim50$, implying that the gas 
is shocked, and that the shock is strong. 

Under these circumstances, the temperature of the shocked gas could be as high as

\begin{equation}
T_{\rm shock} = \frac{3~\mu{m}}{16k} \:\: V_{\rm rel}^2 \:\:,
\end{equation}

\noindent which is $\sim14\times10^6$~K, or $kT_{\rm shock}\sim1200$~eV. 
This is more than sufficient to account for the early presence of 
coronal lines.

However, photoionization may still have a role in the coronal
line emission. As the SS phase commences, an additional hard 
ionizing source of radiation becomes available (in addition 
to shock heating). The impact of this source is considerable, 
especially in the case of \nher\, since it has been shown to 
have the hottest galactic SS source of all well documented SS sources
\citep[see][]{2014ASPC..490..345P}. As the bottom panel of 
Fig.~\ref{aavso} shows, the degree of ionization for the Si 
ion slowly increases with time as the SS source becomes hotter. 
The strength of the \fion{Si}{x} 1.43\mic\ 
line (IP = 351~eV), the strongest coronal line in the spectrum,
likely reflects the extreme high temperature of the central WD surface. 
The overall picture that emerges is that both collisional and 
photoionization are operational, but the key result from this 
work is that collisional ionization \textit{preceded} photoionization, 
since coronal lines were seen about a week before the SS 
phase was detected. Thus, a problem that is hard 
to resolve, whether collisional or photoionization is 
responsible for coronal emission in CNe, \textit{can be}
resolved as has been done here by correlating 
the temporal sequence in which the X-ray SS phase and the 
NIR coronal emission appear.

\subsection{Temperature of the coronal gas}

The coronal lines were first detected on day~11.51, and likely
the coronal emission was entirely the result
of collisions. We can, therefore, estimate the gas temperature using
the fluxes of the various silicon lines at this time. We estimate the 
relative proportions of Si ions as a function of temperature using
the prescription of \cite{1990ApJ...352..307G}, the ionization equilibrium
data in \cite{arnaud_rothenflug1985} and the effective collisional strength values 
$\Psi$ listed in Table~\ref{coronals}. We find the gas temperature
$T_{\rm gas}\simeq10^{5.57 \pm 0.05}$~K, ($kT_{\rm gas} \sim27$~keV). 
This value is consistent with the upper limit on the flux in the 
\fion{Si}{x} 1.43\mic\ line, which requires that $T_{\rm gas}$ 
be less than $10^{5.7}$~K. This value of $T_{\rm gas}$ is
similar to that found in the shock-excited gas around the 
recurrent nova RS~Oph, in which the ejected material runs 
into, and shocks, the secondary wind \citep{2007ApJ...663L..29E, 2009MNRAS.399..357B}. 
We assume $T_{\rm gas}\simeq10^{5.5}$~K in what follows.

\subsection{Abundance estimate}

\cite{1990ApJ...352..307G, 1993ApJS...88...23G} describe a method for estimating the
abundance ratios for selected species.
For the present we use dereddened fluxes from Table~\ref{coronals} 
for lines where there are no blends. Using the \fion{Al}{v} 2.90\mic\ line,
we find that $n(\mbox{Al})/n(\mbox{Si})=0.17$ using 
the \fion{Si}{vii} line, and $=0.14$ derived from the 
\fion{Si}{vi} line. A value of $n(\mbox{Al})/n(\mbox{Si})=4.5\times10^{-2}$ is
derived from the \fion{Al}{vi} and \fion{Si}{vi}  lines. Hence the mean $n(\mbox{Al})/n(\mbox{Si})$
is $\simeq 0.11.$ Using the \fion{Ca}{viii} line, we get a more
consistent result, viz., $n(\mbox{Ca})/n(\mbox{Si})=2.6\times10^{-3}$
from the \fion{Si}{vi} line, and $=2.2\times10^{-3}$ from the 
\fion{Si}{vii} line. The latter values represent lower limits as collision strengths
were not explicitly computed at the assumed temperature for the \fion{Ca}{viii} line.
The solar values are 0.081 (Al/Si) and 0.062 (Ca/Si) respectively \citep{2021A&A...653A.141A}.
We tentatively conclude that aluminum is near solar abundance, and 
calcium is under abundant relative to silicon, with respect to solar values. A more
detailed abundance determination awaits future modeling efforts.

Synoptic multi-wavelength studies of this remarkable nova 
continue as the system returns to quiescence. 

\begin{acknowledgments}

The authors appreciate extensive scientific discussions with A. Evans.
The authors also thank the anonymous referee for suggestions that improved
the manuscript. This Letter is  based on observations obtained under IRTF program 2021A012,
and 2021B996. It is also based on observations obtained for program 
GN-2021B-FT-101 at the international Gemini 
Observatory, a program of NSF's NOIRLab, which is managed by 
the Association of Universities for Research in Astronomy (AURA)
under a cooperative agreement with the National Science Foundation
on behalf of the Gemini Observatory partnership: the National
Science Foundation (United States), National Research Council
(Canada), Agencia Nacional de Investigaci{\'o}n y Desarrollo
(Chile), Ministerio de Ciencia, Tecnolog{\'i}a e Innovaci{\'o}n 
(Argentina), Minist{\'e}rio da Ci{\^e}ncia, Tecnologia, 
Inova\c c{\~o}es e Comunica\c c\~oes (Brazil), and Korea Astronomy and 
Space Science Institute (Republic of Korea). DPKB is supported by a 
CSIR Emeritus Scientist grant-in-aid, which is being hosted by the 
Physical Research Laboratory, Ahmedabad. KLP acknowledges support 
from the UK Space Agency. CEW acknowledges partial support from
NASA PAST 80NSSC19K0868 that enabled our observations.
We acknowledge with thanks the 
variable star observations from the \textit{AAVSO International Database} 
contributed by observers worldwide and used in this research. 

\end{acknowledgments}


\facilities{IRTF (SpeX, Guidedog), Gemini North (GNIRS), Swift, 
AAVSO, {\it Gaia}}

\software{Figaro, IRAF \citep{1993ASPC...52..173T}, Starlink 
\citep{2014ASPC..485..391C}, Astrophy \citep{2018AJ....156..123A}}


\end{document}